\documentclass{aa}  
\usepackage{graphicx}
\usepackage{txfonts}

\usepackage{lscape}
\usepackage{aas_macros} 
\usepackage{amssymb}
\usepackage{natbib}  
\bibpunct{(}{)}{;}{a}{,}{,} 
\usepackage{longtable}
\usepackage{graphicx}
\usepackage[hang,bf,footnotesize]{subfigure}
\usepackage[figuresright]{rotating}

\def\Sec{${}^{\prime\prime}$\llap{.}}

\begin{document}
\titlerunning{Circumstellar disks in NGC~6611}	
   \title{Correlation between the spatial distribution of circumstellar disks and massive stars in the young open cluster NGC~6611.}

   \subtitle{II: Cluster members selected with Spitzer/IRAC.}

   \author{M. G. Guarcello\inst{1,2} \and G. Micela\inst{2} \and F. Damiani\inst{2} \and G. Peres\inst{1} \and L. Prisinzano\inst{2}     \and S. Sciortino\inst{2}}

   \offprints{mguarce@astropa.unipa.it}

   \institute{Dipartimento di Scienze Fisiche ed Astronomiche, Universit\'a di Palermo, Piazza del Parlamento 1, I-90134 Palermo Italy
   \and
   INAF - Osservatorio Astronomico di Palermo, Piazza del Parlamento 1, 90134 Palermo Italy}

  \date{}

 
  \abstract
   {The observations of the proplyds in the Orion Nebula Cluster, showing clear evidence of ongoing photoevaporation, have provided a clear proof about the role of the externally induced photoevaporation in the evolution of circumstellar disks. NGC~6611 is an open cluster suitable to study disk photoevaporation, thanks to its large population of massive members and of stars with disk. In a previous work, we obtained evidence of the influence of the strong UV field generated by the massive cluster members on the evolution of disks around low-mass Pre-Main Sequence members. That work was based on a multi-band BVIJHK and X-ray catalog purposely compiled to select the cluster members with and without disk.}
   {In this paper we complete the list of candidate cluster members, using data at longer wavelengths obtained with Spitzer/IRAC, and we revisit the issue of the effects of UV radiation on the evolution of disks in NGC~6611.}
   {We select the candidate members with disks of NGC~6611, in a field of view of $33^{\prime}\times34^{\prime}$ centered on the cluster, using IRAC color-color diagrams and suitable reddening-free color indices. Besides, using the X-ray data to select Class~III cluster members, we estimate the disks frequency vs. the intensity of the incident radiation emitted by massive members.}
   {We identify 458 candidate members with circumstellar disks, among which 146 had not been revealed in our previous work. Comparing of the various color indices we used to select the cluster members with disk, we claim that they detect the excesses due to the emission of the same physical region of the disk: the inner rim at the dust sublimation radius. Our new results confirm that UV radiation from massive stars affects the evolution of nearby circumstellar disks.}
   {}


   \maketitle
%

\section{Introduction}
\label{intro}
	Circumstellar disks around young Pre-Main Sequence (PMS) stars have been deeply studied since the discovery of their infrared emission \citep{Mendo66,Mendo68}. Until now, several disk models have been developed and a large number of star-forming regions have been observed, in order to understand the physical properties and the evolution of circumstellar disks. Several authors (e.g. \citealt{Ha01}) claim that the emission from the inner region of the disks, responsible for NIR excess in T-Tauri stars, declines on a timescale between $\sim$1 and $\sim$10 Myr. Important clues about the typical timescale of disk evolution come from the studies of PMS stars that have cleared the inner region of their circumstellar disks, thus showing excesses in mid- and far-infrared but not in near-infrared bands. These stars with disk are usually considered to be in a transitional phase between Class~II and disk-less Class~III PMS stars.\par
The evolution of circumstellar disks can be influenced by nearby massive stars, as it was shown by the observations of the evaporating protoplanetary disks in the Trapezium, near the massive star $\Theta^1$ Ori (see, for example, \citealp{Stor99} and references therein). In this case, the photoevaporation process, induced by UV radiation emitted by $\Theta^1$ Ori, dissipates the nearby disks on timescales shorter than $10^6$ years. On the other hand, several authors (for example \citealp{Eis06}) claim that the externally induced photoevaporation is not an efficient mechanism for the truncation of circumstellar disks around low mass PMS stars. The debate about this topic is still open.\par
	Comparing the spatial variations of disk frequencies respect to the positions of massive stars in young open clusters it is possible to study how the radiation from massive stars affects the evolution of circumstellar disks and the formation of new generations of stars. \citet{Bal07} and \citet{io07}, hereafter GPM07, applied this approach to two young clusters: NGC~2244 and NGC~6611 respectively. In their Spitzer/IRAC-MIPS study of NGC~2244 (distant 1.4-1.7 Kpc from the Sun and with an age of about 2-3 Myr), \citet{Bal07} found that disk frequency drops quickly at a distance of 0.5 parsec from the massive cluster members. For larger distances, however, stars with disk do not experience effects induced by massive stars, since the disk frequency is not spatially correlated with their positions. \par
Also the young open cluster NGC~6611, in the Eagle Nebula, is an ideal target for this kind of study, thanks to its rich PMS population and its large number of massive members (56 with spectral classes earlier than B5, \citealp{Hil93}) that are distributed irregularly in the central region of the cluster. In order to obtain a reliable list of cluster members GPM07 used a membership criterion based on infrared excesses (to select members with disk) and X-ray emission (to select members without disk). In fact, thanks to the high X-ray luminosity of PMs stars, this criterion allows to obtain an unbiased sample of PMS cluster members. With this method, a total of 1122 candidate PMS stars associated to NGC~6611 or to the outer part of the Eagle Nebula were identified. In NGC~6611, GPM07 found evidences that the UV radiation from massive stars have dissipated the disks around nearby PMS members on short timescales, since the disks frequency declines for small distances from massive members. To obtain this result, GPM07 calculated the flux emitted by all the massive stars incident on each cluster members and then obtained the disk frequency for various bins of incident flux. To this calculation, the projected distances were used, instead the real ones. However, GPM07 showed that this approximation have not affected their result. \par
	The effects of the energetic radiation from massive stars on the evolution of nearby circumstellar disks are also studied with simulations of the evolution of open clusters. For example \citet{Ada06} claim that externally induced photoevaporation does no affect the evolution of disks in clusters with less than 1000 members. Instead, more populated clusters, as shown by \citet{Fatu08}, can be environments in which disks are efficiently evaporated in small timescales, since they have intense UV fields practically independent from the number of stellar members. \par
This paper is the follow up of the GPM07 study, enriching the list of cluster members by using Spitzer/IRAC data and confirming their principal results; both works, therefore, will be the base of a subsequent paper, focused on the analysis of the Spectral Energy Distributions of selected cluster members. \par
	The Spitzer/IRAC observations of NGC~6611 used in this work are obtained from the Galactic Legacy Infrared Mid-Plane Survey Extraordinaire (GLIMPSE; \citealt{Ben03}). The {\it Infrared Array Camera} (IRAC), mounted on Spitzer Space Telescope \citep{Faz04}, allows studying stars with circumstellar disk in four infrared bands, centered on 3.6~$\mu m$, 4.5~$\mu m$, 5.8~$\mu m$ and 8.0~$\mu m$. In this region of the spectrum, the contribution from the stellar photosphere is usually small compared to the emission from disk and envelope, and the effects of extinction by interstellar medium is smaller than at shorter wavelengths. These facts permit a more reliably selection of stars with circumstellar disks than using 2MASS data alone. \par
Our paper is organized as follow: in Sect. \ref{catal} we show the results of the cross identification between GLIMPSE catalog and that compiled in GPM07; in Sect. \ref{colcolsp} and \ref{Qpar} we will describe the identification of stars with disk respectively using IRAC color-color diagram and suitable reddening-free color indices that combine optical and NIR colors; in Sect. \ref{compat} we will compare the two diagnostics used in this paper; in Sect. \ref{memcat} we describe the catalog of the cluster members; in Sect. \ref{final} we will re-examine the results obtained in GPM07 using the updated member list.\par

\section{Multi-wavelength catalog}
\label{catal}

	\subsection{Description of GPM07 multiband catalog and NGC~6611 parameters}
	\label{gpmcat}

GPM07 compiled a multi band catalog of NGC~6611, using optical observations in BVI bands in a $33^{\prime}\times34^{\prime}$ field of view centered on the cluster, obtained with ESO 2.2m/WFI camera (PI: Mundt), 2MASS public data of the same sky region and an X-ray observation performed with Chandra/ACIS-I in a $17^{\prime}\times17^{\prime}$ region (obs.ID 978; \citealt{Lin07}). Hereafter, we will call WFI FOV and ACIS FOV the fields of view of the respective instruments; the latter FOV is contained in the former and it is approximately at its center. The catalog compiled in GPM07 consists of 38995 sources falling in the WFI FOV. \par
	Using this catalog, GPM07 estimated the age of the PMS members (that are mostly younger than 3 Myr), the distance of the cluster ($\sim$ 1750 parsec), the anomalous reddening law (with $R_V\sim3.3$), the average extinction for cluster members ($A_V=2.6^m$) and the relaxation time for the core (4.2 Myr, longer than the age of PMS members). The age of the cluster, however, is not well constrained as other cluster parameters. For example, several authors claimed that star formation in this nebula started about 6 Myears ago (see \citealp{Gvara08} for a detailed discussion about the age spread in this cluster). \par
In the ACIS FOV we found 997 X-ray sources using {\it PWDetect}, a wavelet-based source detection algorithm \citep{Dami97}, with 10 expected spurious sources. Among the X-ray sources with optical counterpart, 31 have $V$ magnitudes and $V-I$ colors compatible with foreground main sequence stars. This fraction of X-ray sources ($\sim 3\%$) is comparable to the estimated contamination in similar X-ray observations (see, for example, \citealt{Dami06}). However, we will not exclude a priori these sources from our list of candidate cluster members, since they can be Class~II YSOs with optical colors altered by effects due to their PMS nature (as disk gas accretion on stellar surface). Instead, they will be marked with a specific tag.  \par 


	\subsection{Cross identification with GLIMPSE catalog}
	\label{glimcat}
 
 	\begin{table*}[ht]
	\centering
	\caption {Results of the cross-identification.}
	\vspace{0.5cm}
	\begin{tabular}{ccccc}
	\hline
	\hline
	Number of stars& WFI detection & 2MASS detection & IRAC detection & Number of X-ray sources \\
	\hline
	$146  $    &$no$&	$no$&	$no $&	$146$\\
	$20476$    &$no$&	$no$&	$yes$&	$0  $\\
	$2732 $    &$no$&	$yes$&	$no$&	$26 $\\
	$17768$    &$no$&	$yes$&	$yes$&	$64 $\\
	$12535$    &$yes$&	$no$&	$no$&	$196$\\
	$1071 $    &$yes$&	$no$&	$yes$&	$178$\\
	$2648 $    &$yes$&	$yes$&	$no$&	$74 $\\
	$2782 $    &$yes$&	$yes$&	$yes$&	$370$\\
	\hline
	\hline
	\multicolumn{5}{l} {} 
	\end{tabular}
	\label{cross}
	\end{table*}

	GLIMPSE observations in the field containing NGC~6611 cover the entire $33^{\prime}\times34^{\prime}$ WFI FOV, with the exception of a region at North-West of about  $7^{\prime}\times10^{\prime}$. Following the explanatory manual\footnote{available at http://www.astro.wisc.edu/sirtf/docs.html}, we selected 41985 IRAC sources falling in this region. \par
We identified the sources in common to both the GLIMPSE catalog and the catalog compiled by GPM07 with an identification radius of 0\Sec3; this value is equal to the astrometric precision of GLIMPSE catalog relative to 2MASS, as reported in the explanatory manual (in GPM07 we used the 2MASS Point Source Catalog as astrometric reference). In this work we also include 146 X-ray sources without any optical or 2MASS counterpart, which were not included in GPM07, to identify their possible IRAC counterpart.  With the chosen identification radius, we expect about 26 spurious identifications, evaluated as in \citet{Dami06}.\par
	The results of the cross-identification are summarized in Table \ref{cross}. We found 2782 stars with WFI, 2MASS and IRAC detections. This sample shows a significant spatial clumping in correspondence of the cluster and it includes 370 X-ray sources, suggesting that it is mostly composed by cluster members. \par
\begin{figure}[]
	\centering	
	\includegraphics[width=9cm]{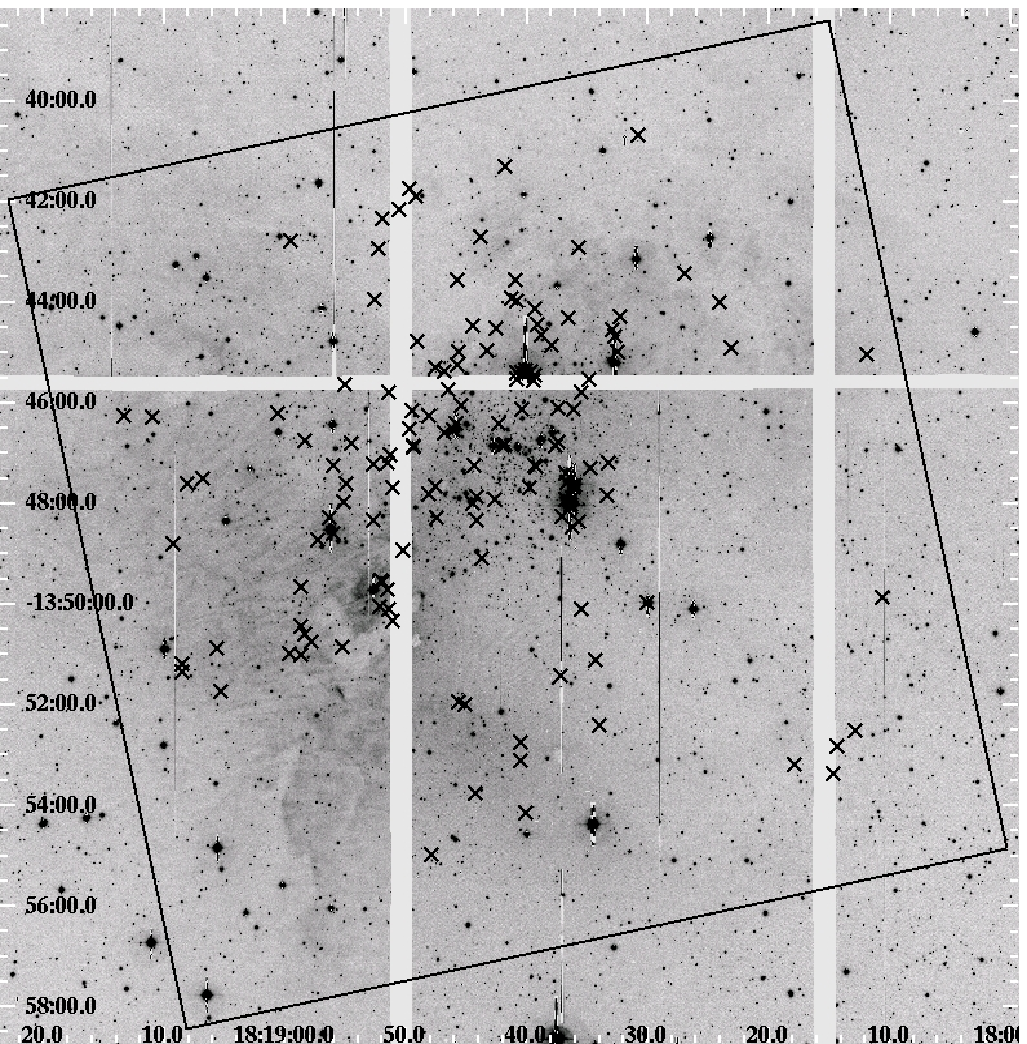}
	\caption{WFI image, in the I band, of the central region of NGC~6611. Crosses identify the X-ray sources without both optical and infrared counterpart. The box outlines the $17^{\prime}\times17^{\prime}$ ACIS FOV.}
	\label{xmiss}
	\end{figure}

	
Table \ref{cross} also shows that the 146 X-ray sources without any optical or 2MASS counterpart are not detected even in IRAC observations. Fig. \ref{xmiss} shows the spatial distribution of these sources, evidently clustered near the center of the ACIS FOV. They can be low-mass cluster members, with masses lower than the limit of our optical-IR data, since our catalog includes X-ray sources down to the limiting magnitudes in WFI, 2MASS and IRAC observations. However, it is also possible that some of these objects can be extragalactic sources, detectable in X-ray since the nebula is less dense in correspondence of the cavity cleared by the cluster itself.
	
\begin{figure}[]
	\centering	
	\includegraphics[width=9cm]{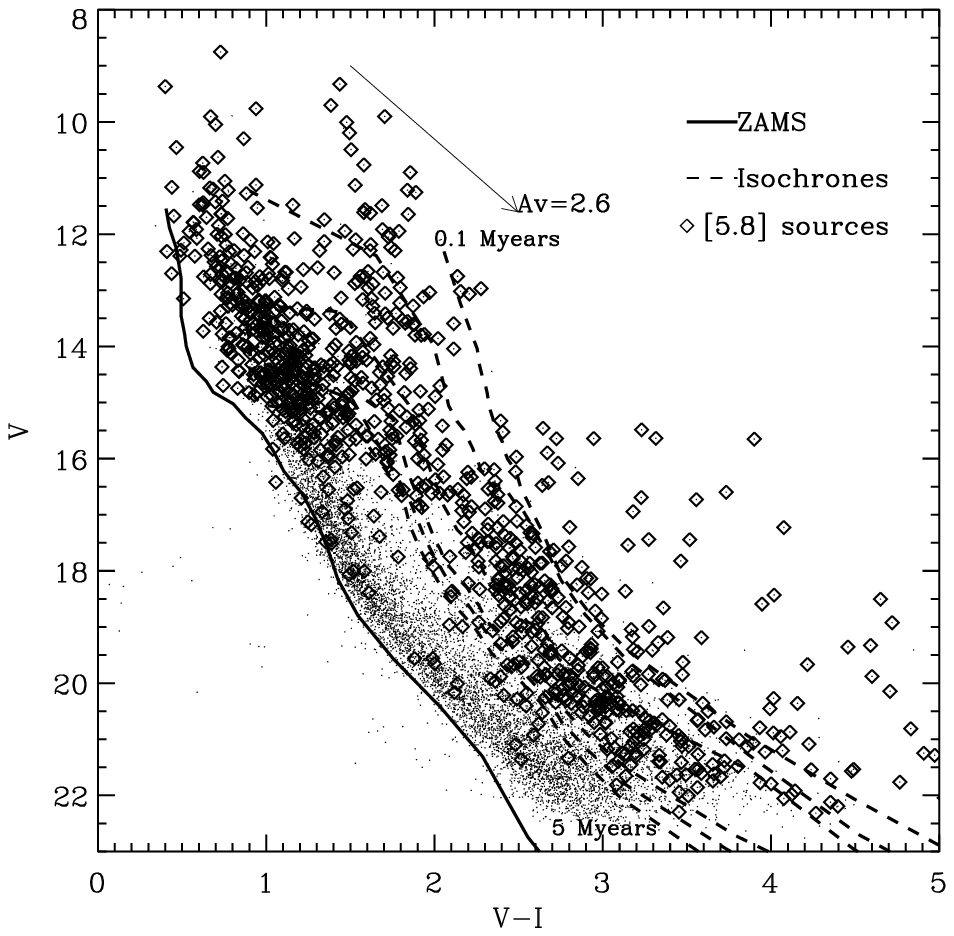}
	\caption{$V$ vs. $V-I$ diagram of stars in WFI FOV. The thick solid line is the ZAMS (from \citealt{Sie00}), at the distance of 1750~pc and with $A_V=1.45^m$ (the average extinction of field stars, evaluated in GPM07). The dashed lines are the isochrones at 0.1, 0.25, 1, 2.5, 3 and 5 Myears, with the average extinction appropriate for cluster members ($A_V=2.6^m$). The extinction vector is obtained from the law of \citet{Muna96}. The diamonds are optical sources with emission at 5.8 $\mu m$. }
	\label{vvi3}
	\end{figure}

GPM07 used the detection in ACIS observation as membership criterion. This was justified by the fact that almost all our X-ray sources with optical emission are PMS stars compatible with an age between 0.1 and 3 Myears. Fig. \ref{vvi3} shows the $V$ vs. $V-I$ diagram for the stars in WFI FOV with errors in $V$ smaller than 0.1$^m$ and in $V-I$ smaller than 0.15$^m$. In this diagram, diamonds mark the optical sources detected also at 5.8$\mu m$. About 54\% of these objects are bright stars ($V \leq 16^m$), while the remaining 46\% are fainter and mainly concentrated in the PMS region of the color-magnitude diagram. This region overlaps with that traced by X-ray sources, confirming the youth of these optical-IRAC stars. The $V$ vs. $V-I$ diagrams of the sources detected in the other IRAC bands have similar characteristics. Hereafter, the color-magnitude and color-color diagrams of this paper will include only stars with errors in the specific magnitude smaller than 0.1$^m$ and in color smaller than 0.15$^m$.
	
	In the next two sections we will describe the criteria adopted to select stars with circumstellar disk based on the excesses in IRAC colors, and we will compare them with the criterion based on 2MASS data, used in GPM07. These criteria are applied to stars with magnitude errors smaller than $0.1^m$; this is a very stringent rule if applied to faint stars: at 3.6$\mu m$ more than 50\% of stars fainter than $12.5^m$ have error greater than $0.1^m$. The same 50\% fraction is found at $12^m$ at 4.5$\mu m$ and at $10.5^m$ at both 5.8$\mu m$ and 8.0$\mu m$. It is evident, then, that a criterion for the selection of stars with disk using all IRAC bands simultaneously does not allow to select faint sources with disk, since most of them have large errors in the IRAC bands at longer wavelenghts. We will partially overcome this problem using two independent disk diagnostics: the usual color-color IRAC diagram (Sect. \ref{colcolsp}) and a diagnostic based on suitable reddening-free color indices (Sect. \ref{Qpar}), involving optical+2MASS photometry and only a subset of IRAC bands.  


\section{T-Tauri stars from IRAC color-color diagram}
\label{colcolsp}

	The [3.6]-[4.5] vs. [5.8]-[8.0] diagram is an excellent tool to identify stars with circumstellar disk (see, for example, \citealt{Alle04}). In this diagram stars with photospheric colors are clustered around the origin, while stars with disk show colors significantly different from zero. The two populations can be easily distinguished even if the stars are affected by large interstellar extinctions. This is possible since the reddening vector is almost vertical (due to a very small reddening in [5.8]-[8.0]) or it points to bluer [5.8]-[8.0] colors (due to the partial overlap of the IRAC 8.0 $\mu m$ band with the interstellar silicate feature), depending on the adopted extinction law; see, for example, \citet{Mege04} and \citet{Fla07}. Unlike what happens for other possible infrared diagrams, then, in the IRAC color-color plane the locus of reddened photospheres and that of sources with intrinsic red colors are not blended, even for large extinctions.\par
With this diagram it is also possible to distinguish roughly embedded Class~I and Class~II T-Tauri YSOs (Young Stellar Objects): the former have both colors redder than the latter since the presence of the collapsing envelope surrounding the star and the disk; see, for example, \citet{Alle04} who used the models developed by \citet{dale98,dale99,dale01} for Class~II stars and by \citet{Ken93} and \citet{Calve94} for embedded Class~I stars.  \par
	 
	\begin{figure}[]
	\centering	
	\includegraphics[width=9cm]{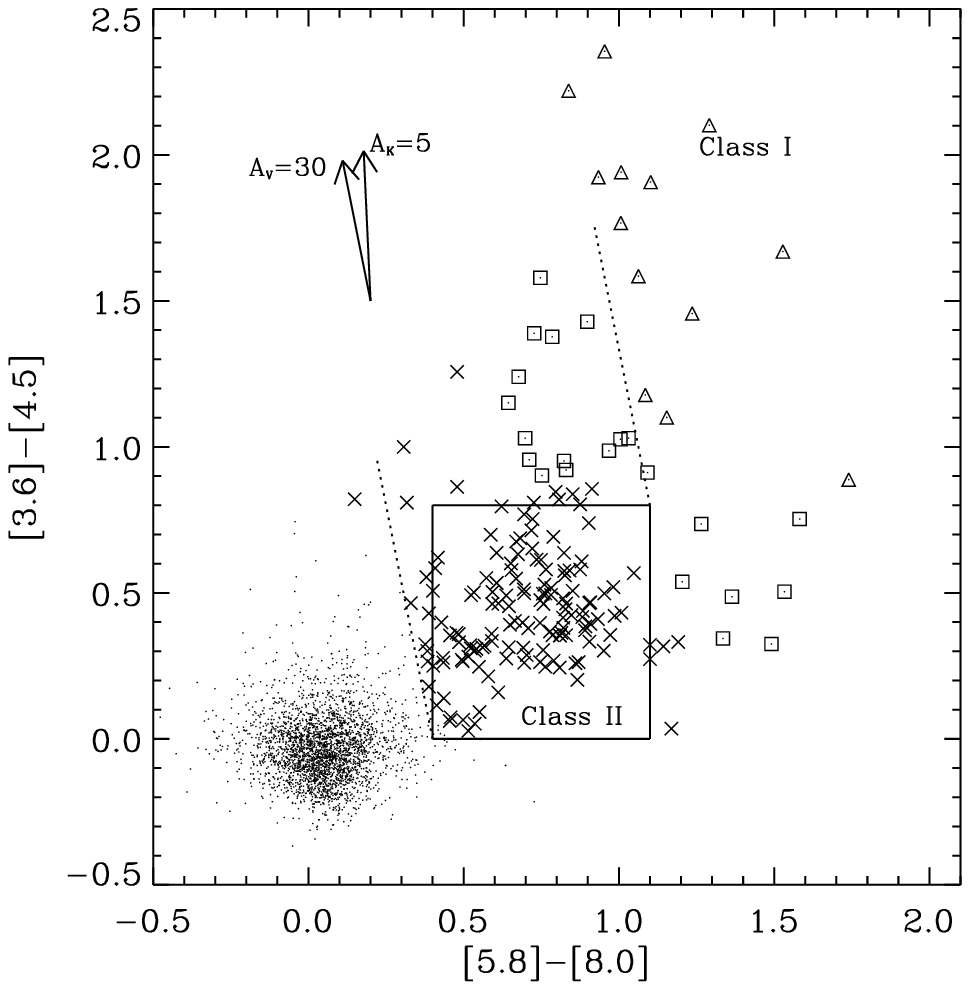}
	\caption{Color-color diagram for IRAC sources in WFI FOV (dots). The box marks the approximate locus of Class~II YSOs (crosses). The dashed lines roughly separate reddened photospheres (left), reddened Class~II YSOs (center) and Class~I YSOs (in the upper right part of the diagram, marked with triangles). Squares mark stars either classifiable as Class~II or Class~I YSOs. The reddening vectors with A$_V$=30$^m$ and A$_K$=5$^m$ are obtained from \citet{Mege04}, and \citet{Fla07}, respectively.}
	\label{CCsp}
	\end{figure}

	\subsection{Classification of selected YSOs}
	\label{ysoclass}

GLIMPSE data of the Eagle Nebula were already analyzed, in combination with 2MASS data, by \citet{In07}. We reanalyze the data, using also different methods, for several reasons. First, we note that the selection of stars with $K$ excess performed by \citet{In07} and GPM07 is different. The latter used the reddening free color indices, a method much more efficient than the use of the T-Tauri locus \citep{Mey97} in infrared color-color diagrams, adopted by the former. Moreover, in GPM07 only stars with color errors ($\sigma_{colors}$) smaller than 0.15$^m$ have been used, and we apply the same approach here in order to be more conservative as possible. With this condition we produce a different IRAC color-color diagram respect to \citet{In07}. In the present paper, this diagram is used together with reddening-free color indices defined with IRAC bands, in order to perform a selection of the cluster members with disk as much as possible model-independent. Furthermore, 2MASS and IRAC data are complemented by optical and X-ray data, that are fundamental to asses the nature of the candidate members. \par
	Fig. \ref{CCsp} shows the color-color diagram for the IRAC sources in WFI FOV used to classify the YSOs. Stars with colors typical of photospheres are clustered around the origin, with a spread consistent with photometric uncertainties and reddening, and are clearly separated from sources with intrinsic red colors. \par
To select and classify stars with disk, we used the reddening vector obtained by \citet{Mege04} from the reddening law of \citet{Mat90}, shown as the inclined vector in Fig. \ref{CCsp}. The problem of the correct extinction law in IRAC bands is still open. For this reason, in Fig. \ref{CCsp} we also show the extinction vector (the $A_K=5^m$ arrow) obtained from the mean reddening law recently estimated by \citet{Fla07} in the direction of five nearby star-forming regions. It is evident that the difference between the two extinction laws is relevant only for a very large extinction.\par
	Using this diagram, we identify 147 Class~II YSOs as the stars within the box in Fig. \ref{CCsp} (crosses), taking into account photometric uncertainties and extinction. This box was obtained from the models of stars with circumstellar disk developed by \citet{dale98,dale99,dale01}. In addition, 13 stars are unambiguously classified as Class~I YSOs (triangles in Fig. \ref{CCsp}). \par
22 stars (marked with squares in Fig. \ref{CCsp}) are sources with only one IRAC color redder than the Class~II locus. Those with $[3.6]-[4.5]\geq0.8^m$ can be either Class~I sources or very embedded Class~II YSOs; those with $[5.8]-[8.0]\geq1.2^m$ are usually interpreted as Class~II YSOs, with no emission detected from the inner region of the disk, due to a larger inner hole or to the disk inclination with respect to the line of sight. However, as suggested by \citet{Ken93}, also Class~I YSOs with large centrifugal radii\footnote{the distance from the central star at which the infalling material from the envelope with most angular momentum, i.e. near the equatorial plane, hits the circumstellar disk.} and optically thin envelopes, that give a silicate band in emission, may have a spectral energy distribution compatible with the latter colors. For these reasons, all these 22 stars are not clearly classified. The possible contamination by extragalactic sources is discussed in the next section.\par
\begin{figure}[!h]
	\centering	
	\includegraphics[width=7cm]{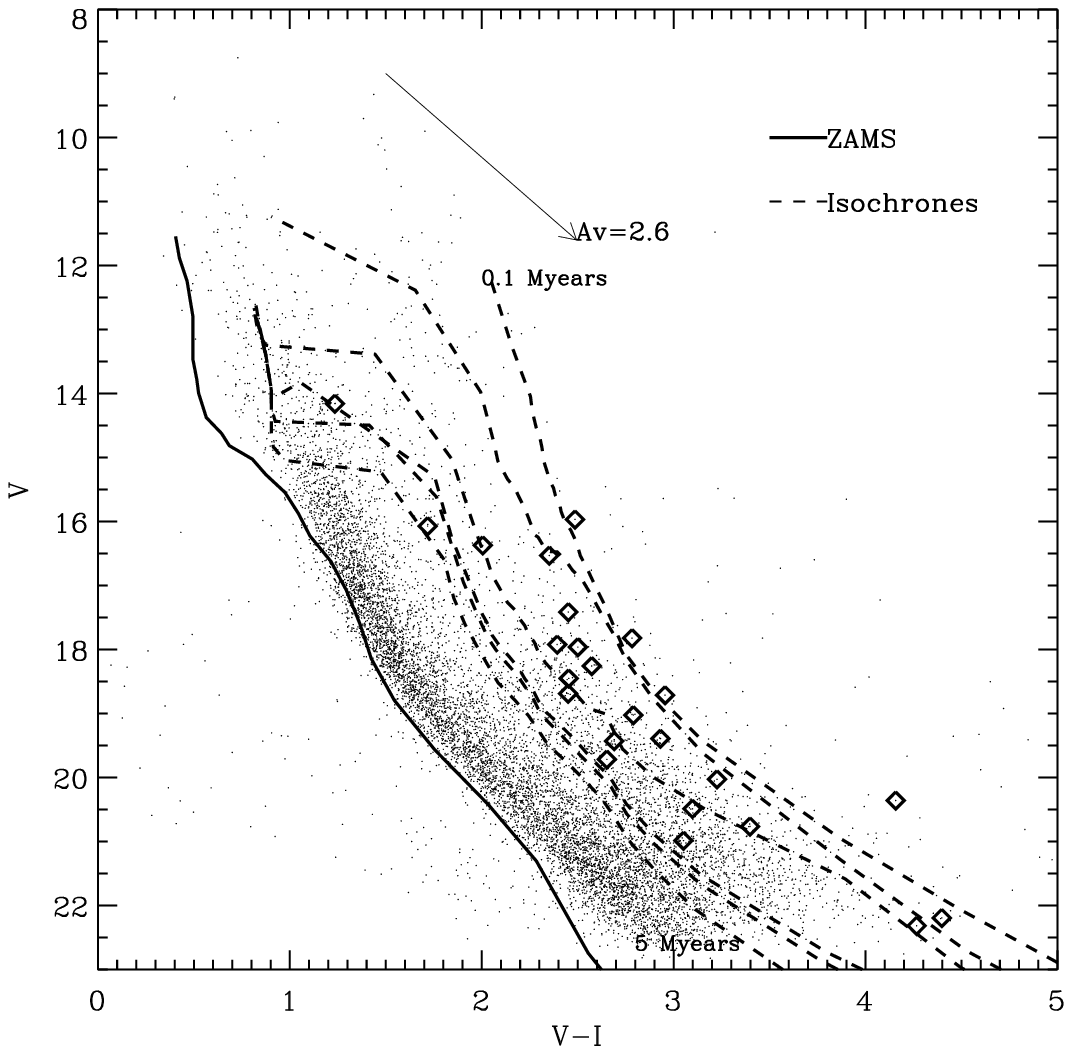}
	\caption{$V$ vs. $V-I$ diagram analogous to that in Fig. \ref{vvi3}. Diamonds mark candidate stars with disk selected by the IRAC color-color diagrams but not in GPM07}
	\label{vvisp}
	\end{figure}

	The IRAC color-color diagram allows us to select a total of 182 members with excesses, with 120 new identifications with respect to GPM07, where BVIJHK bands were used. This result is not in disagreement with GPM07, since in that work it was not possible to classify most of these stars. Among these 120 stars, in fact, only 48 have good measurements in $K$, 25 in both $V$ and $I$ (shown in Fig. \ref{vvisp}) and just 9 in all these three bands simultaneously. We also note that 19 of these stars are also X-ray sources (among the total of 57 that are inside the ACIS FOV) and that the $V-I$ and $V$ values of the subsample with good WFI photometry are compatible with the cluster. \par
\begin{figure}[!h]
	\centering	
	\includegraphics[width=7cm]{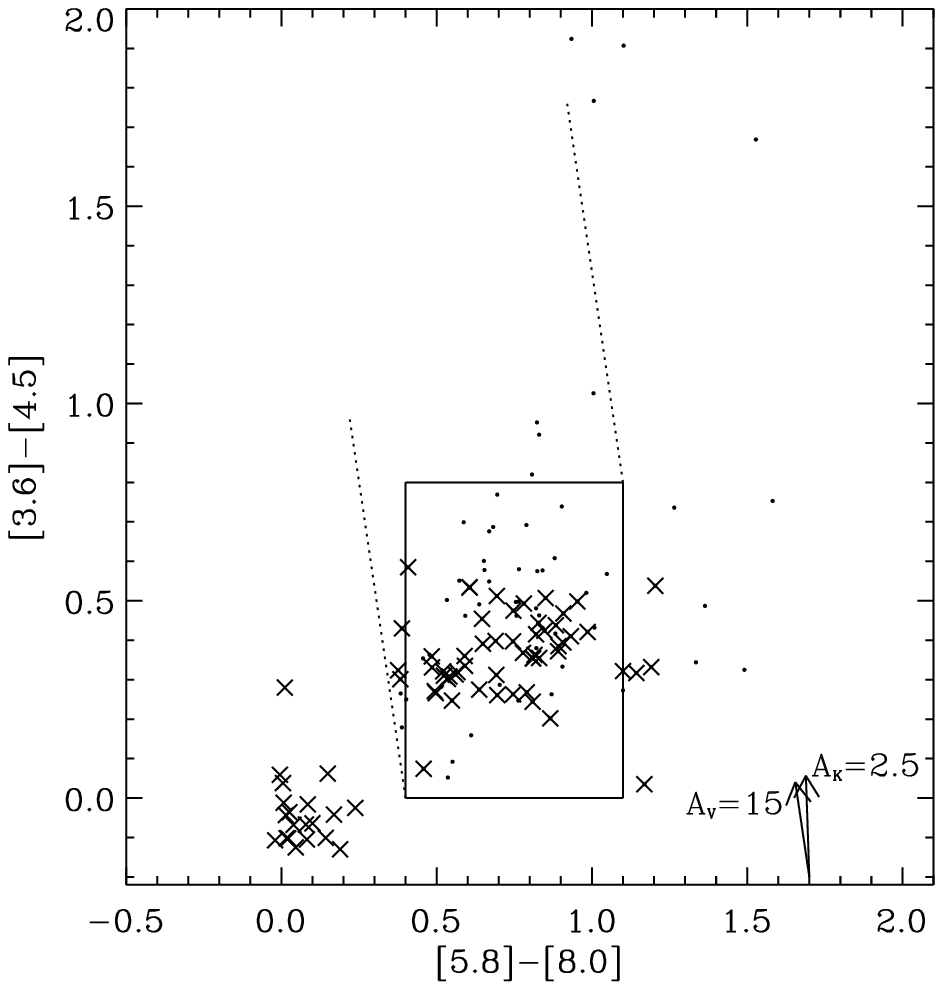}
	\caption{IRAC color-color diagrams of the X-ray sources (crosses). Dots mark the stars with disk selected from the IRAC diagram of Fig. \ref{CCsp}, for the sky region falling in the ACIS FOV. The Class~II locus and the reddening vectors are analogous to those in Fig. \ref{CCsp}.}
	\label{CCsp2m}
	\end{figure}

 
	Fig. \ref{CCsp2m} shows the IRAC colors of 76 X-ray sources with good photometry in the four IRAC bands. A large number of these sources (55) are classified as Class~II YSOs while none as Class~I sources. This is likely due to lower luminosity in X-ray of  Class~I objects respect to more evolved YSOs, in accord to that showed by previous studies about star-forming regions analyzed with X-ray observations deeper than our one (i.e. \citealt{Lore08}).
	
\subsection{Contaminating sources}
\label{cont}

In order to evaluate the contamination due to extragalactic sources, we use the criteria defined in \citet{Gute08}. These criteria allow to identify candidate AGNs, or galaxies dominated by PAH emission, using various color-color and color-magnitude IRAC diagrams. Using these criteria, we identify only 4 stars that can be extragalactic sources, but their $A_V$ values predicted by the optical colors are too small (between 2.5$^m$ and 5$^m$) to confirm this classification.\par
	However, in IRAC color-color plane PAH-rich (Polycyclic Aromatic Hydrocarbons) star-forming galaxies can be found in the locus at $-0.1^m<[3.6]-[4.5]<0.6^m$ and $[5.8]-[8.0] > 1^m$ (in the part of the diagram redder than the Class~II locus). We have selected 25 candidate disk-bearing members in this region of the diagram. Some of them could be PAH-rich star-forming galaxies, and we will definitively classify them in our subsequent paper.\par 

	\subsection{Spatial distribution of Class~II and Class~I YSOs}
	\label{classdis}

	\begin{figure}[]
	\centering	
	\includegraphics[angle=0,width=9cm]{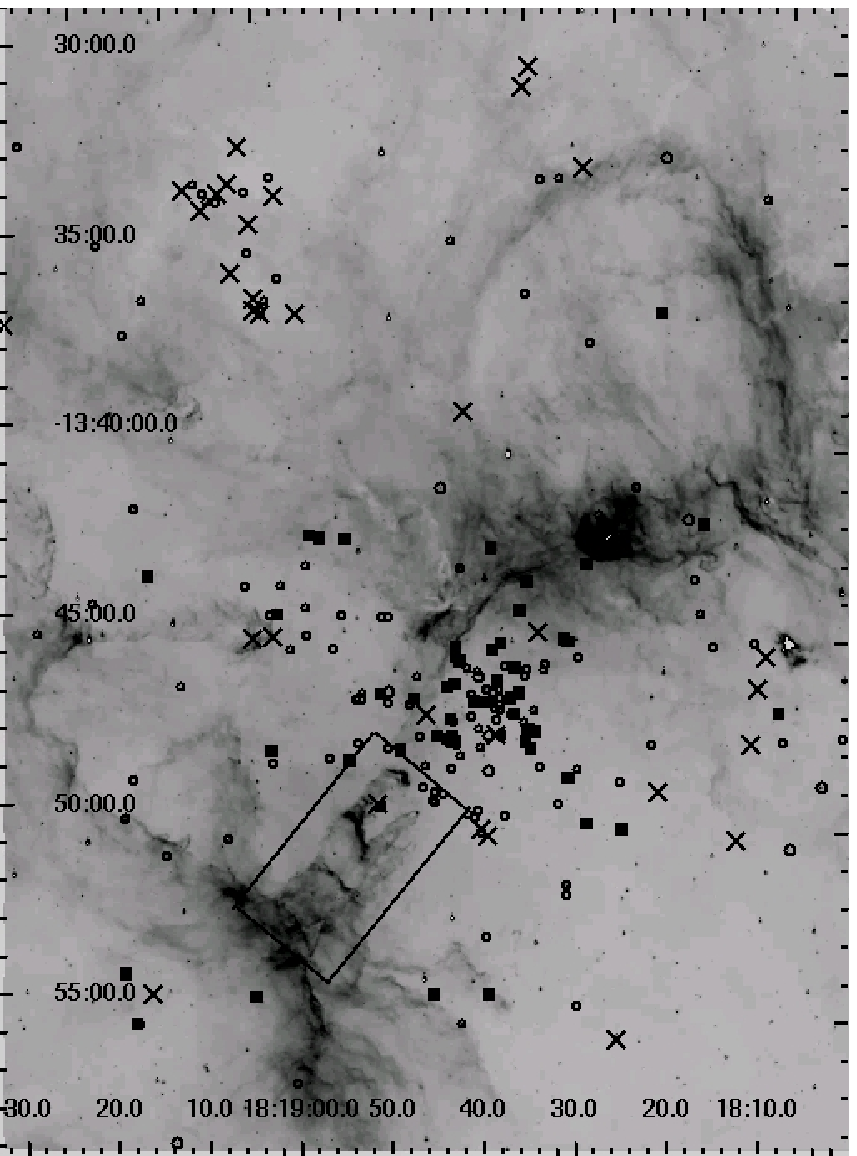}
	\caption{IRAC image of the Eagle Nebula at 8.0 $\mu m$, centered on the cluster and covering approximately a region of $22^{\prime} \times 32^{\prime}$. The circles mark Class~II YSOs, while the crosses mark Class~I YSOs and stars compatible with both classifications. The small boxes mark massive stars. The box delimits the structures known as ``elephant trunks''. North is up, East is to the left.}
	\label{classspadis}
	\end{figure}

Fig. \ref{classspadis} shows the IRAC image of the Eagle Nebula at 8.0$\mu m$, with over plotted the Class~II objects (circles) and the sources sorted either as Class~I or as reddened Class~II YSOs (crosses). In the center of the cluster (approximately in the center of the image) there is a lack of sources of the latter group, with respect to the outer regions. This is not a real result, since it is due to the intense diffuse emission of the dense structures in the central region of the nebula, like the elephant trunks inside the box in Fig. \ref {classspadis}, that complicates the extraction of IRAC point sources. These structures, in fact, are heated directly by the radiation from the massive members of NGC~6611, and are very bright in the IRAC bands at longer wavelengths. Besides, these diffuse structures are very dense, accounting for a very high extinction inside them. \par
	The evidence that star formation activity is still ongoing in the central region of NGC~6611 was provided by several authors. For example, \citet{Mcc02}, using observations at high spatial resolution, showed that the trunks are stellar ``nurseries'', with several embedded YSOs inside them. The trunks are continuously eroded by the intense incident UV flux emitted by nearby massive cluster members. As results, some of the young sources formed inside the trunks are emerging from the photodissociation regions that delimit these structures. As effect of the intense diffuse emission by the nebula, we identify only one Class~I YSO emerging from the trunks, at $\alpha=18:18:52$ and $\delta=-13:49:38$ (the cross inside the box in Fig. \ref{classspadis}). \par 
As shown in Fig. \ref{classspadis}, 12 sources classified either as Class~I or reddened Class~II are clustered in a region at North-West. Only one of the few Class~II stars present in this region has a faint optical counterpart (with $V=23.34^m$). Therefore, this small cluster can be associated to a denser intra cluster medium and/or to more recent star formation events, if compared with the center of NGC~6611. The presence of this rich star-formation site, already suggested by \citet{In07}, shows that the star formation activity is still ongoing in the whole nebula, and not only in the central region. \par 

\section{Cluster members from reddening-free color indices}
\label{Qpar}

In Sect. \ref{catal}, we discussed the different sensitivities in the four IRAC bands and that this can limit the selection of stars with disk with the IRAC color-color diagram. To partially overcome this problem, we select the stars with infrared excesses also using four suitable reddening-free color indices, that will be described in detail in the following (see also GPM07 and \citealt{Dami06}). \par

	\subsection{Properties of the $Q$ color indices}
	\label{Qprop}

	\begin{figure*}[]
	\centering	
	\includegraphics[width=19cm]{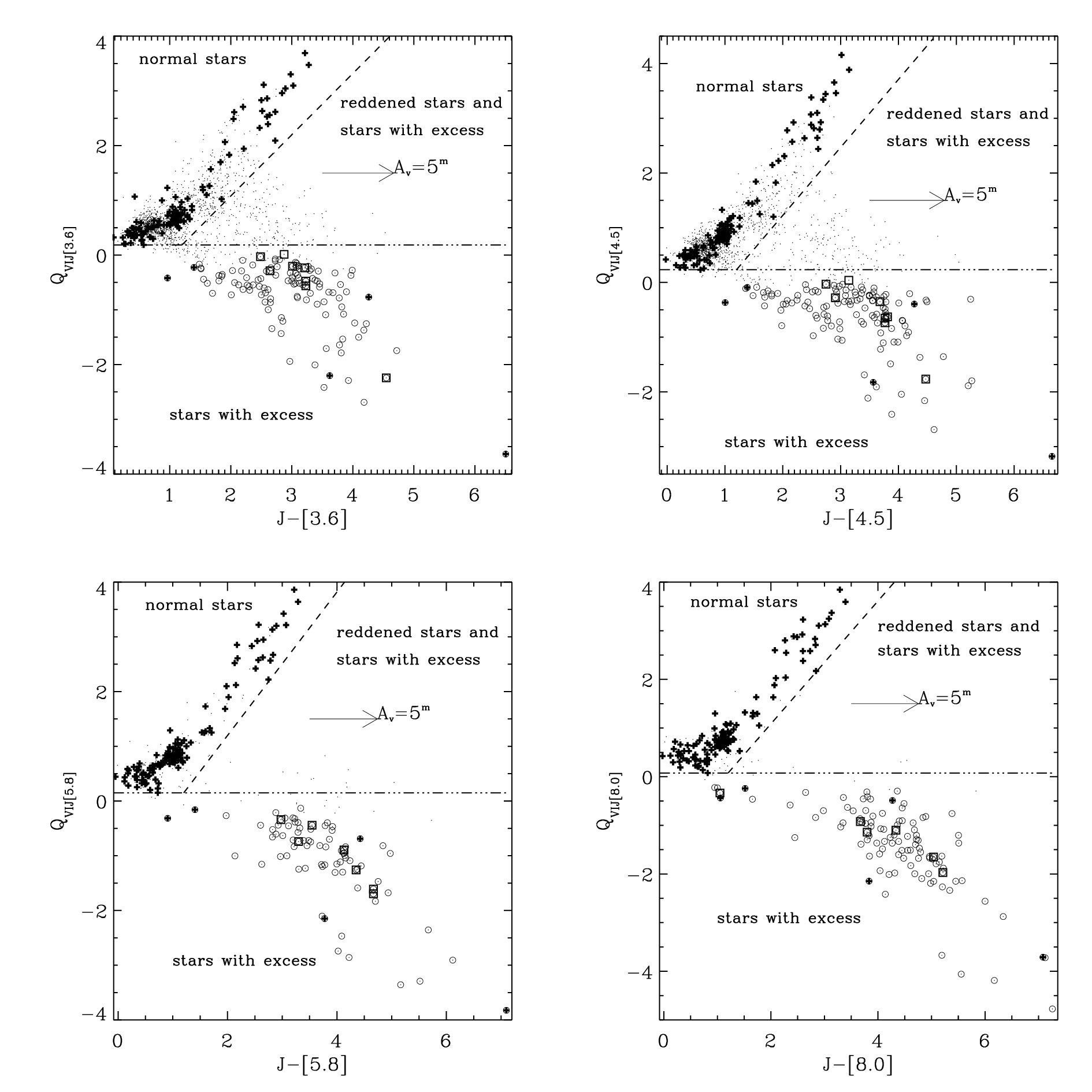}
	\caption{Diagrams $Q_{VIJ[sp]}$ vs. the $J-[sp]$ colors for the stars in WFI FOV (points). Circles mark stars with excess in the corresponding index; the horizontal dotted-dashed lines are the lower limits for photospheric indices; the inclined dashed lines separate the locus of normal stars from that of reddened sources (or stars with small excesses). Crosses mark the optical sources used to define these limits (as explained in the text). The squares mark stars with multiple identifications.}
	\label{Qdiag}
	\end{figure*}

$Q$ indices are defined in order to compare infrared colors with $V-I$, the latter assumed to be representative of photospheric emission:
	
	\begin{equation}
	Q_{VIJ[sp]} = \left( V-I \right) - \left( J-[sp] \right) \times E_{V-I}/E_{J-[sp]}
	\label{Qdef}
	\end{equation}
	
where $Q_{VIJ[sp]}$ is the index; $V$, $I$ and $J$ are the magnitudes in the respective bands; $[sp]$ is the magnitude in a specific IRAC band; $E_{V-I}$ and $E_{J-[sp]}$ are the reddening in the corresponding colors. $E_{V-I}$ has been obtained from the reddening law of \citet{Muna96}, while the extinction in IRAC bands has been taken from the reddening law of \citealt{Mat90} (see Fig. \ref{CCsp}). The effectiveness of this approach is that $Q_{VIJ[sp]}$ indices allow us to determine excesses in stars for which the magnitudes in all the IRAC bands are not well measured simultaneously, if $V$, $I$ and $J$ magnitudes are known. \par
	The identification of stars with excesses is performed using the diagrams $Q_{VIJ[sp]}$ vs. $J-[sp]$ in Figs. \ref{Qdiag}. In these diagrams the extinction gives a shift only along the $x-axis$, while the excess in the $[sp]$ band downward the $y-axis$. In this way, stars with excesses and reddened photospheres can be separated, even if a significant number of stars remain unclassified between reddened photospheres and stars with excesses. In Figs. \ref{Qdiag} the stars with excesses are marked by circles; they are defined as the stars with the $Q_{VIJ[sp]}$ indices significantly (i.e. by more than 3 times the error in $Q$ indices) smaller than the horizontal dotted-dashed lines, that mark the lowest limits of photospheric $Q_{VIJ[sp]}$ indices. We define these limits as the low boundaries of the loci, in these diagrams, of the IRAC sources with optical counterpart that are clustered around the origin in the IRAC color-color plane in Fig. \ref{CCsp}. The spatial distribution and the positions in the other color-color and color-magnitude diagrams of these stars confirm that they are field normal stars or cluster members without disk. In Fig. \ref{Qdiag} they are marked with crosses, and it is evident that they are separated either from the stars with excesses and the reddened sources. Hereafter we will call {\it NS locus} that of the normal stars in the diagrams in Fig. \ref{Qdiag}; {\it EX locus} that of the stars with excesses; {\it UNC locus} that of the stars compatible with both interpretations. \par
The minimum excesses necessary so that a star with disk can be selected, by the use of the $Q_{VIJ[sp]}$ here and $Q_{2MASS}$ in GPM07, is mass and age dependent. To analyze these property of $Q$ indices, we compute these minimum excesses in [3.6] and $K$ for stars with masses equal to 0.5, 1, 1.5, 2, 2.5, 3, 4, 5 solar masses and with ages of 0.1, 0.5, 1, 2.5, 3.5, 5 Myears. Their {\it photospheric} indices are computed with the colors predicted by the evolutionary tracks of \citet{Sie00}, using the color transformation of \citet{KH95}. The minimum excesses are then the excesses in [3.6] or $K$ necessary so that their $Q$ indices become more negative than the lower limit of photospheric colors for more than 3$\sigma$, where $\sigma$ is the mean error in $Q$ of the sources in the WFI FOV. \par 	

	\begin{figure*}[]
	\centering	
	\includegraphics[width=11cm]{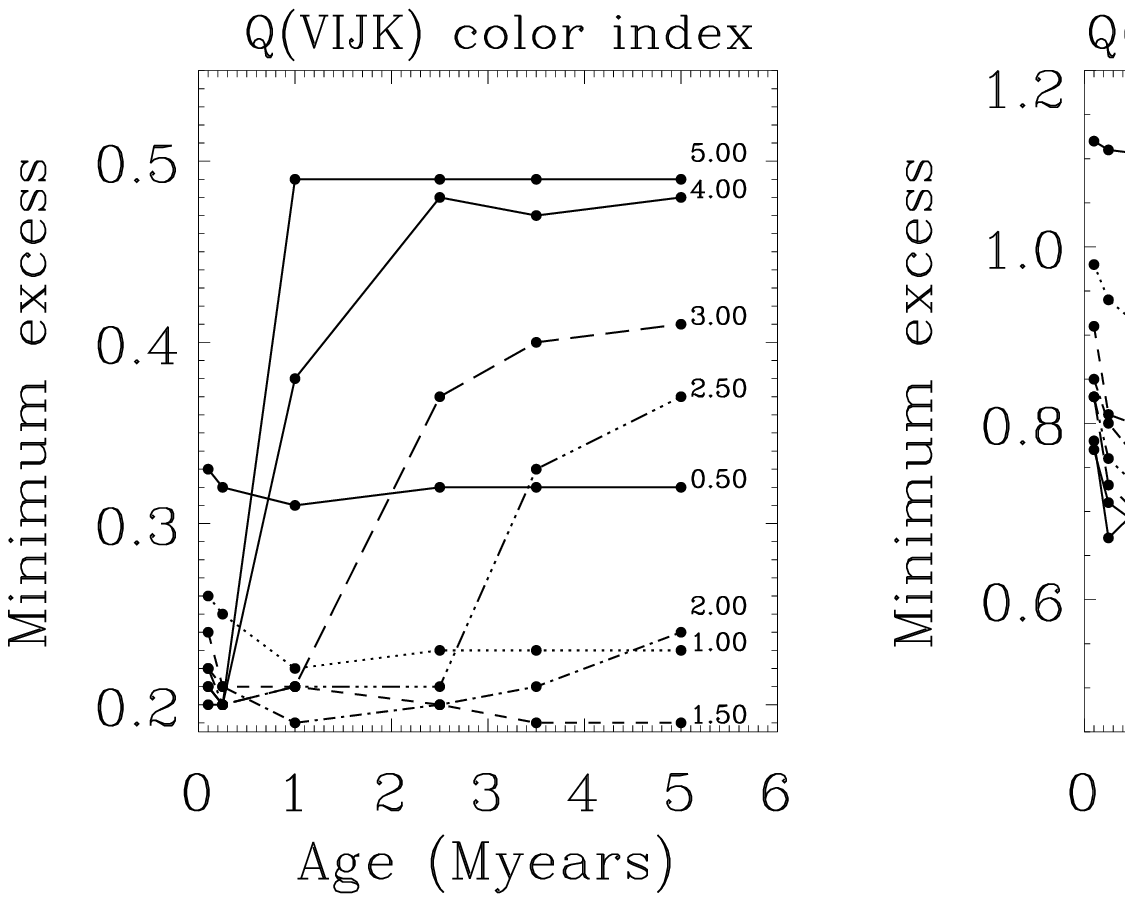}
	\caption{Trend of the minimum excesses necessary to identify stars with excesses in $K$ and [3.6] for the studied masses and ages values. The excesses in $K$ are detected with the use of $Q_{VIJK}$ index, defined in GPM07; the excess in [3.6] with $Q_{VIJ[3.6]}$.}
	\label{Qtrend}
	\end{figure*}

	Fig. \ref{Qtrend} shows the variation of the minimum excesses for one among the four indices used in GPM07 to detect excesses in $K$ and for that used here to detect excesses in [3.6]. It is evident that for each mass and for each $Q$ index the minimum excesses have irregular paths with increasing age. This is due to the irregular behavior of the indices when both the involved colors become more red (or more blue), as in comparing normal stars with different effective temperatures. \par
Fig. \ref{Qtrend} also shows that $Q_{VIJK}$ is more efficient in detecting excesses in stars with lower masses respect to $Q_{VIJ[3.6]}$ index. This behavior depends on the shape of the isochrones in the $Q$ diagrams. For example, the minimum of the 5Myears isochrone in $Q_{VIJ[3.6]}$ of Fig. \ref{Qdiag} occurs at $\sim$B9 spectral type, corresponding to 2.5 $M_{\odot}$. This implies that stars of lower mass with small excesses easily fall in the $UNC$ region, where we are unable to distinguish between small excesses and reddening and therefore we need stronger excesses to unambiguously asses the presence of disks. The corresponding minimum in $Q_{VIJK}$ diagram occurs at $\sim$K4 spectral type (using the 5 Myears isochrone), making this index much more efficient in detecting excesses in stars with smaller masses.


	\subsection{Stars with $Q$ excesses}
	\label{Qres}

	\begin{table}[]
	\centering
	\caption {The first column yields the $Q_{VIJ[sp]}$ indices used in this work; the second column shows the number of stars with excess detected with each $Q_{VIJ[sp]}$ index; the third column shows the total number of stars for which we compute the indices.}
	\vspace{0.5cm}
	\begin{tabular}{ccc}
	\hline 
	\hline
	$Q$ index& Number of stars with excess & Number of stars \\
	\hline
	$Q_{VIJ[3.6]}$	&113    &1746\\
	$Q_{VIJ[4.5]}$	&124    &1474\\
	$Q_{VIJ[5.8]}$	& 79    & 357  \\
	$Q_{VIJ[8.0]}$	&104    & 297  \\
	\hline
	\hline
	\multicolumn{3}{l} {A total of 174 stars with excesses are selected} 
	\end{tabular}
	\label{comQ}
	\end{table}

	Table \ref{comQ} gives the total number of stars for which we compute the indices and the number of stars with excesses. These numbers can be lower limits since the criterion we adopt to detect the stars with excesses is very conservative. \par
If we consider the total number of stars in our catalog with good photometry in the bands used to define the indices (third column in Table \ref{comQ}), we find a higher percentage of stars with excesses in the IRAC bands at longer wavelengths. This confirms that a large number of optical sources are detected in the IRAC bands with longest wavelengths thanks to their nature of PMS stars, as suggested in Sect. \ref{glimcat}. \par
	Table \ref{comQ} also shows a drop of the numbers of selected stars with excesses at [5.8], due to the smaller numbers of stars with good measurement in this band. This is an effect of the decrease in sensitivity with increasing wavelength. Except for 2 stars, the excesses in the three short IRAC wavelength bands are  correlated. This suggests that they are related to the same physical region of the disk. This region can only be the inner rim at which disk dusts sublimate. As proposed by several authors (i.e. \citealt{dale98,dale99,dale01}), in fact, this region is optically thick and it emits as a blackbody at the dust sublimation temperature (between 1500K and 2000K). 16 stars have excess only at [8.0]. This is a consequence of the fact that [8.0] can be affected by the  10$\mu m$ silicate emission feature. A Class~II YSO, then, can show excess in [8.0] even if the inner disk is mostly evacuated, as we can hypothesize for at least 4 among the 16 members of NGC~6611 with excesses detected only with $Q_{VIJ[8.0]}$ (since the other 12 fall in the {\it UNC loci} in the other $Q_{VIJ[sp]}$ diagrams).

	\section{Comparison among the used disk diagnostics}
	\label{compat}
	
	Using the $Q_{VIJ[sp]}$ indices, we select 30 new stars with excesses (including 5 X-ray sources) not found with the $Q_{2MASS}$ indices in GPM07. Among these stars, as explained, 16 have excesses only at [8.0]; the other 14 have large errors in $K$ band or fall in the {\it UNC} loci of the $Q_{2MASS}$ diagrams in GPM07. \par
Similarly, 186 stars with excesses in 2MASS bands (mostly in K) selected in GPM07, are not classified as stars with excesses with $Q_{VIJ[sp]}$ indices. This number of stars can seem very large, since usually more stars with excesses are observed at longer wavelengths. However, it must be noted that 94 of these stars have poor IRAC photometry, consistently with the different sensitivities between 2MASS and GLIMPSE catalogs, and therefore they are not included in IRAC sample. All the other 92 fall in the {\it UNC} loci of the diagrams in Fig. \ref{Qdiag}. \par
	We conclude that the set of indices defined in this work is consistent with that defined in GPM07 (with 144 sources with at least one excess in 2MASS bands and at least one in IRAC bands). This consistency means that if $Q$ indices detect excesses in some particular 2MASS or IRAC band, the lack of excesses in the other bands is almost always due to poor photometry or to the ambiguity between excesses and reddening, as explained above. This is compatible with a scenario in which excesses in $K$ (and $H$) are due to the same physical region of the disk of the excesses in IRAC bands. It is important to note, however, that this consistency rules only between the $Q$ indices, that are an efficient diagnostic for the selection of Class~II YSOs having moderate disk inclination (with respect to the line of sight), because of the use of the optical and $J$ bands. The color-color IRAC diagram is instead a peerless tool for the selection of embedded YSOs and Class~II stars with highly inclined disk. Therefore the two methods may be considered complementary. \par

	\begin{figure}[]
	\centering	
	\includegraphics[width=8cm]{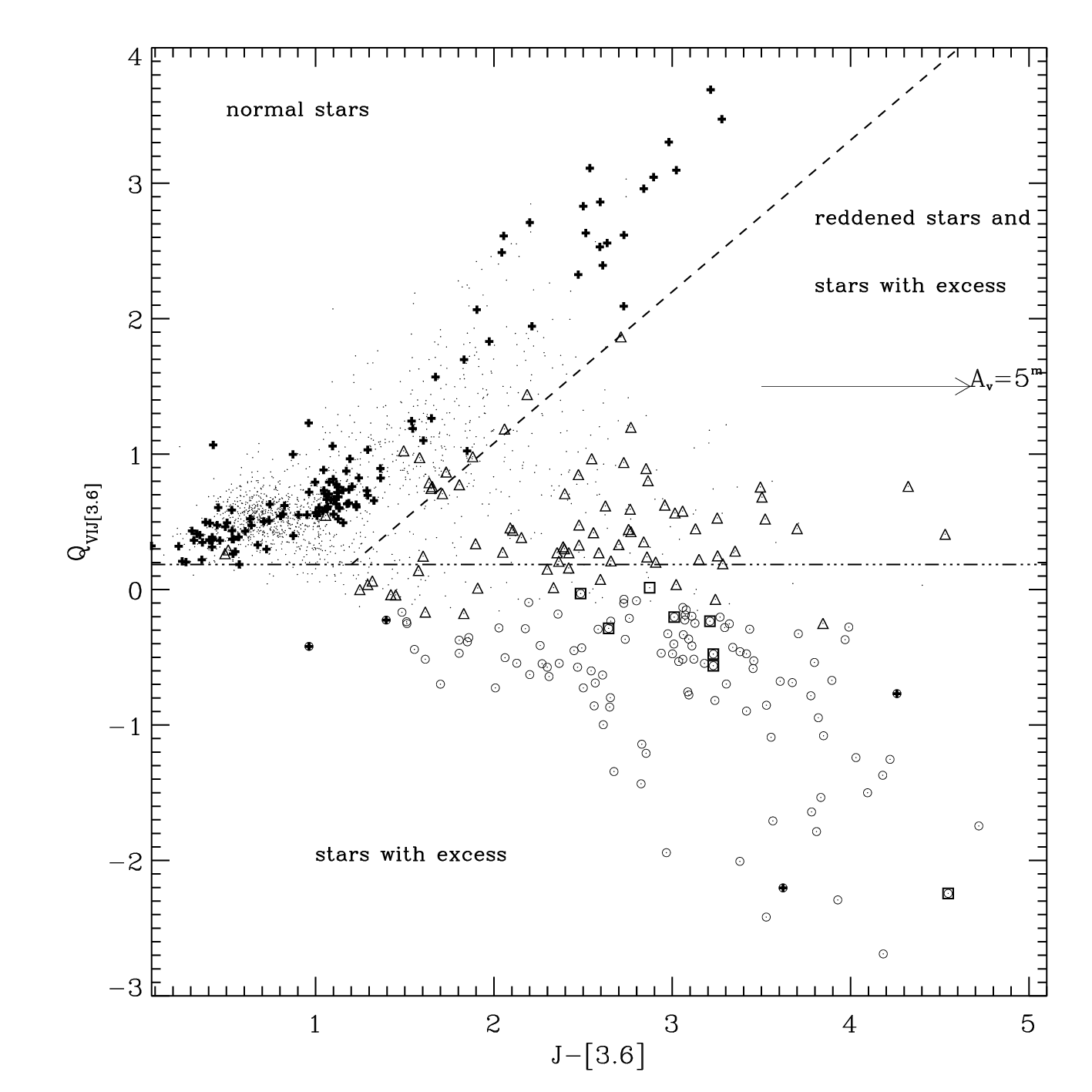}
	\caption{Diagram of $Q_{VIJ[3.6]}$ vs. $J-[3.6]$, similar to the diagrams in Fig. \ref{Qdiag}; the stars with excesses detected in 2MASS but not in IRAC bands are marked as triangles.}
	\label{Q1}
	\end{figure}

\begin{figure}[!h]
	\centering	
	\includegraphics[width=6cm]{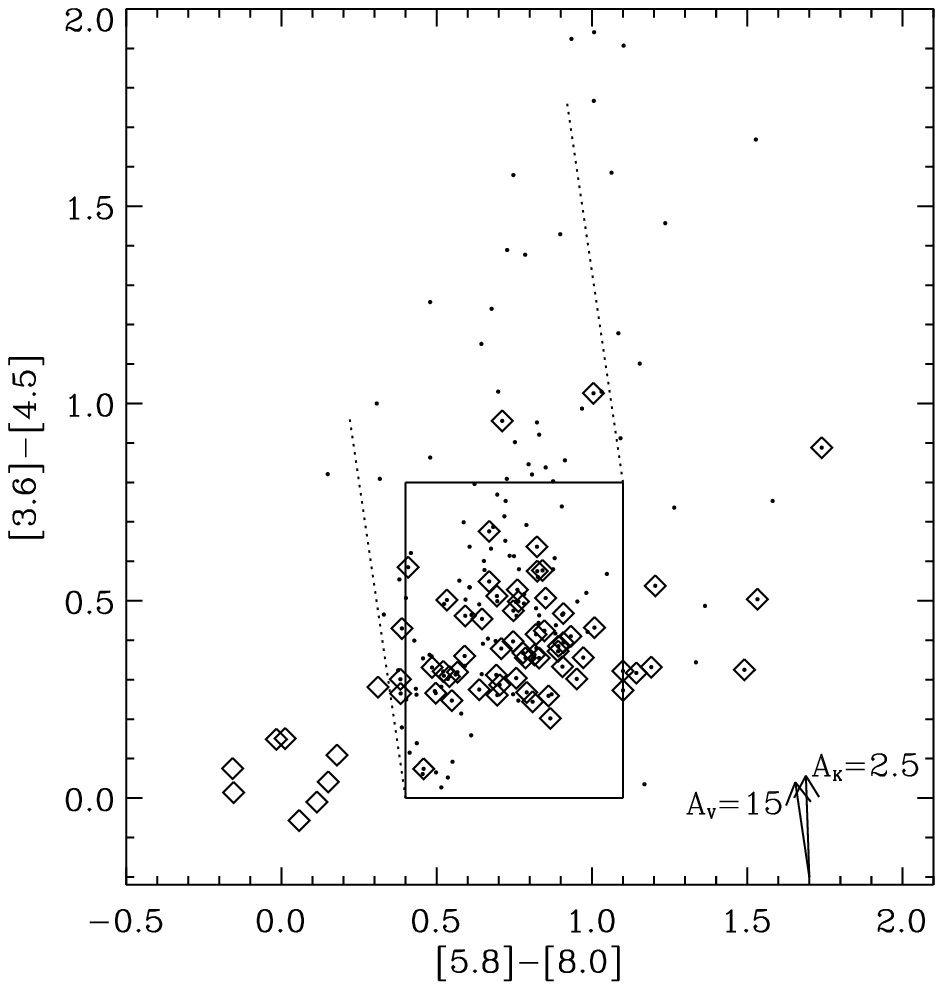}
	\caption{IRAC color-color diagrams of the stars with disk selected with the IRAC diagram of Fig. \ref{CCsp} (points) and of the stars with excesses in 2MASS bands selected in GPM07 (diamonds). The Class~II locus and the reddening vectors are analogous to those in Fig. \ref{CCsp}.}
	\label{CCspX}
	\end{figure}

\begin{figure}[!h]
	\centering	
	\includegraphics[width=8cm]{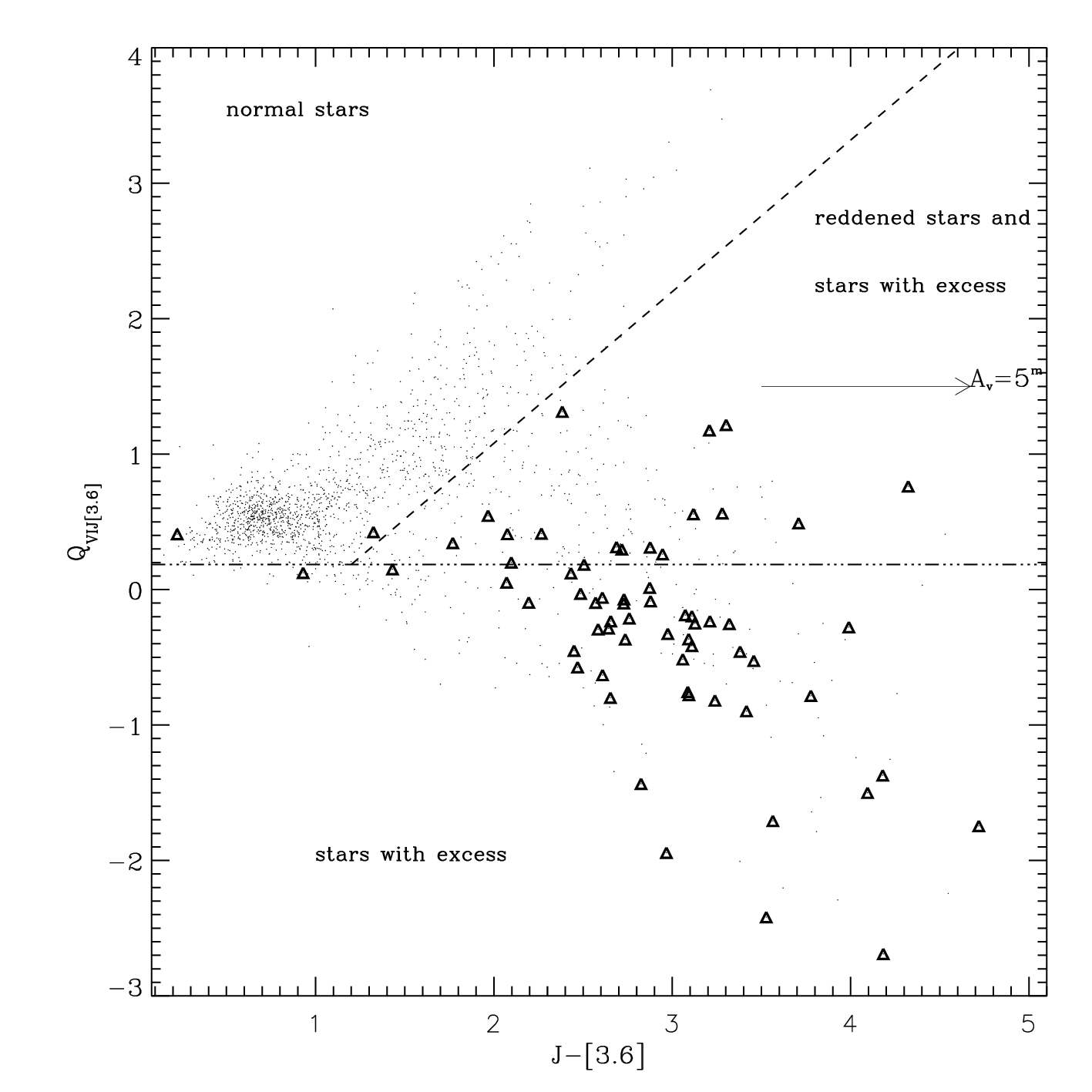}
	\caption{Diagram $Q_{VIJ[3.6]}$ vs. $J-[3.6]$ for the stars in the WFI FOV, with overplotted the stars with excesses selected with the IRAC color-color diagram (triangles).}
	\label{Q1ccsp}
	\end{figure}

Fig. \ref{CCspX} and Fig. \ref{Q1ccsp} show how the excesses detected with the IRAC color-color diagram translate into those detected with $Q$ indices, and vice versa. In Fig. \ref{CCspX} diamonds mark the 70 candidate stars with disk selected in GPM07 with $Q_{2MASS}$ indices having good photometry in all IRAC bands, while dots mark the candidate stars with disk selected with the IRAC color-color diagram. Only 8 sources with excesses detected with $Q_{2MASS}$ indices are close to the origin of the diagram (in the locus of the normal stars), while the other have red IRAC colors. The characteristics of these 8 sources will be studied by SED analysis in our subsequent paper. Fig. \ref{Q1ccsp} shows the $Q_{VIJ[3.6]}$ index of candidate cluster members with disk selected with the IRAC color-color diagram (marked with triangles). As expected, almost no one of these stars fall in the {\it NS locus}. \par
\begin{figure}[]
	\centering	
	\includegraphics[width=8cm]{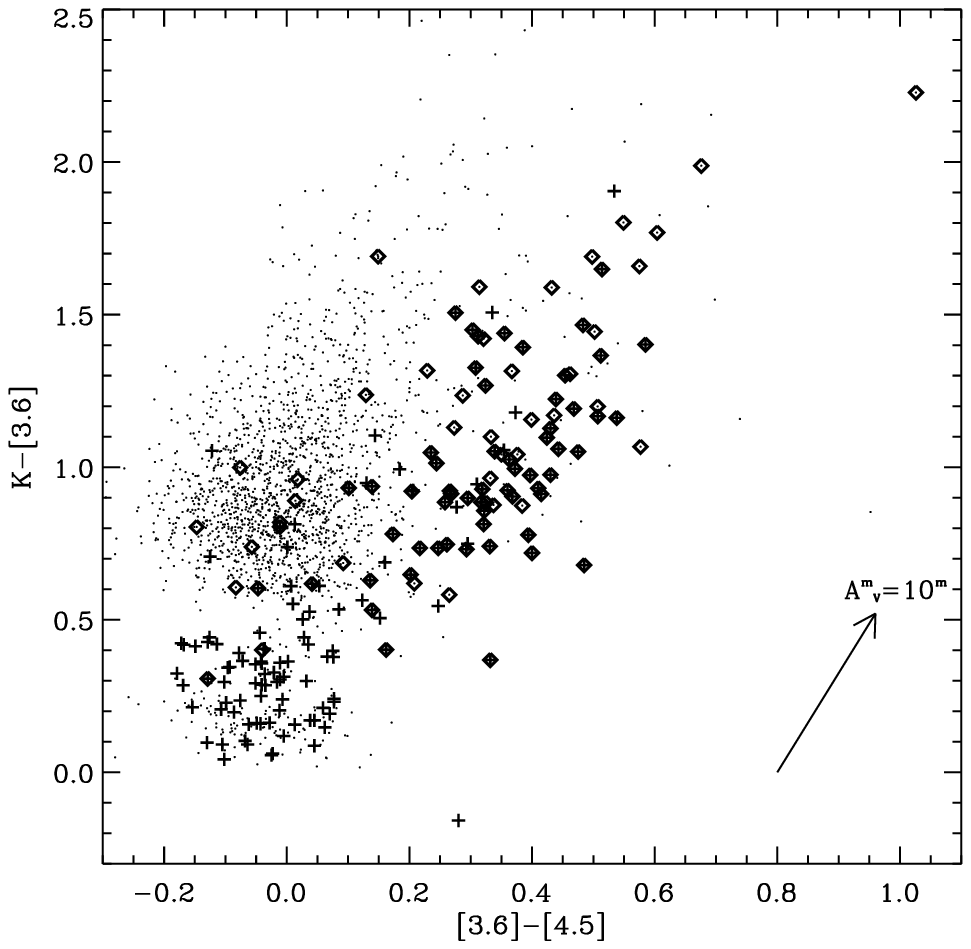}
	\caption{Color-color diagram with $K$ and IRAC bands for the stars in ACIS FOV (points). Diamonds are stars with excesses in 2MASS bands detected with $Q_{2MASS}$ indices defined in GMP07, while crosses mark X-ray sources. The reddening vector, corresponding to $A_{V} = 10^m$, has been obtained from the extinction law of \citet{Mege04}.}
	\label{K112}
	\end{figure}

As further test, Fig. \ref{K112} shows the $K-[3.6]$ vs. $[3.6]-[4.5]$ diagram for the stars in ACIS FOV, with the X-ray sources and the stars with excesses in 2MASS bands selected in GPM07. All the stars with excesses in 2MASS bands have $K-[3.6] \geq 0.7^m$ and/or $[3.6]-[4.5] \geq 0.2^m$, with the exception of a couple of stars that have also both IRAC colors $\sim0$. \par
	In the diagram in Fig. \ref{K112} it is also clear that the distribution of the stars with normal colors (marked with points) shows a gap at $K-[3.6]\sim0.5^m$, likely due to a rapid increase of the interstellar extinction at the distance of the Eagle Nebula: the foreground contaminating sources are clustered around the origin of the diagram and they are clearly separated from the sample of the more extincted stars, dominated by background sources. In fact, the gap between these two samples is populated by a large number of X-ray sources, mostly associated with the nebula.

	\begin{table}[]
	\centering
	\caption {Number of stars with excesses detected with the different diagnostics used in this paper and in GPM07.}
	\vspace{0.5cm}
	\begin{tabular}{cc}
	\hline
	\hline
	Disk diagnostics& Numbers of stars with excesses \\
	\hline
	IRAC color-color diagram	&182\\
	$Q_{2MASS}$ indices	&330\\
	$Q_{VIJ[sp]}$ indices		&174\\
	\hline
	\hline
	\multicolumn{2}{l} {A total of 458 different candidate members with disk are selected.} 
	\end{tabular}
	\label{dd}
	\end{table}

Table \ref{dd} summarizes the number of stars with excesses selected with the diagnostics used here and in GPM07.

	\section{Catalog of the candidate cluster members}
	\label{memcat}

Combining all the diagnostics used in this paper and in GPM07, we select a total of 1264 candidate members of NGC~6611: 790 candidate Class~III members (inside the ACIS FOV) and 474 candidate Class~II and Class~I PMS members (inside the WFI FOV).  \par
	The catalog, available in electronic format, comprises:
\begin{itemize}
\item stars IDs and coordinates;
\item the magnitudes and the errors in BVIJHK and IRAC bands; if some value is not available it is set equal to ``NAN'';
\item a tag ({\it tagx}) that is equal to $1$ if the star is also an X-ray source, otherwise the tag is equal to $0$;
\item a tag ({\it tagM}) that is equal to $A01$ if the star is classified as a candidate members with disk only by its position in the IRAC color-color diagram, $A10$ if the infrared excesses are detected only by $Q$ indices and $A11$ if both diagnostics detect the emission from the disk; $B$ if the star is a disk-less X-ray source; $C$ if it is an X-ray source with optical colors compatible with a foreground main sequence star (see Sect. \ref{catal}).
\end{itemize} 

\section{Spatial variation of disks frequency}
\label{final}
	
	In order to confirm the results obtained in GPM07, we take advantage of the new list of cluster members obtained here. Inside the central $17^{\prime}\times17^{\prime}$ ACIS FOV, where X-ray data are available, we select 790 candidate PMS members without disk, thanks to their X-ray emission and the presence of an optical/IR counterpart, among which 31 are likely foreground contaminants. In addition, in the same FOV we found 257 stars with NIR excesses, 54 more than in GPM07, that are candidate to have a circumstellar disk (118 are also X-ray sources). \par
The average disk frequency inside the ACIS FOV is $ 24\% \pm 2\%$, larger than the value obtained in GPM07 ($19\% \pm 1\%$). \citet{Oli05} found a disk frequency equal to $\sim 58\%$ in a small region inside the ACIS FOV. The different fraction are likely due to the different sensitivities of the two surveys. In fact, the $L$ survey used in \citet{Oli05} is deeper than the GLIMPSE catalog (about 1 magnitude respect to [3.6] sources catalog). Moreover, \citet{Oli05} shown that the disk frequency increases with decreasing mass of the central star, likely since the more massive is the central star the more rapid is the erosion of the circumstellar disk, as it was found in other star forming regions (for example, the study of \citealp{Carpe06}, on the Scorpius OB association). This may be the origin of the discrepancy with \citet{Oli05} results.\par
	However, more than the absolute value of disk frequency, we are interested to its spatial variation respect to the position of massive members, that allows to understand if UV radiation emitted by the latter members may alter the disk lifetimes of nearby T-Tauri YSOs. In order to reach this goal we need only to be sure to use a consistent criterion in different sky positions. GPM07 already used this approach, and we verify here their results using the new list of cluster members. \par
\begin{figure}[]
	\centering	
	\includegraphics[width=6cm]{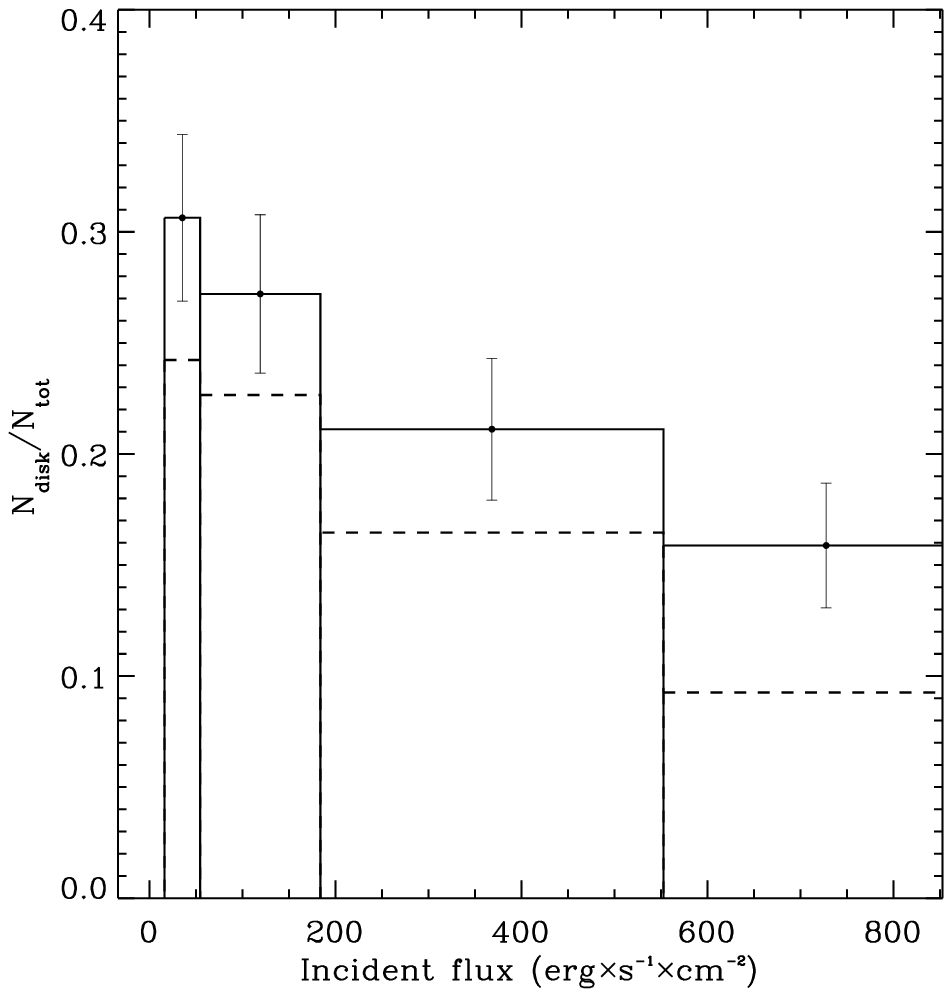}
	\caption{Percentage of members with disk in the ACIS FOV vs. the estimated incident flux from OB members. The four values are $31\% \pm 4\%$, $27\% \pm 4\%$, $21\% \pm 3\%$ and finally $16\% \pm 3\%$, from left to the right.}
	\label{isto}
	\end{figure}

For every cluster member, with and without disk, we compute the incident flux emitted by 52 massive members of the cluster, with spectral class earlier than B5 \citep{Hil93}. To this calculation, we use the projected distances from massive stars, as already explained (Sect. \ref{intro}). We then compute the disk frequencies for various bins of incident flux, as shown in Fig. \ref{isto}. It is evident that the main GPM07 result, i.e. that members with disk are more frequent at larger distances from massive stars, where they are irradiated by lower UV fluxes, is confirmed. Disk frequency, in fact, increases with decreasing incident fluxes: from $31\% \pm 4\%$ in the bin with lowest flux, to  $27\% \pm 4\%$, $21\% \pm 3\%$ and finally $16\% \pm 3\%$ in the bin with highest flux. In the histogram in Fig. \ref{isto}, the bins size are defined in order to have the same number of disk less members in each bin. These disk frequencies are obviously more reliable than those computed in GPM07, thanks to the new and more complete list of cluster members. This paper, then, reinforces the result of GPM07 about the influence of massive stars on the evolution of nearby members with disk in NGC~6611.

	\subsection{Spatial distribution of the cluster}
	\label{cluspadis}

	\begin{figure*}[]
	\centering	
	\includegraphics[width=6.7cm]{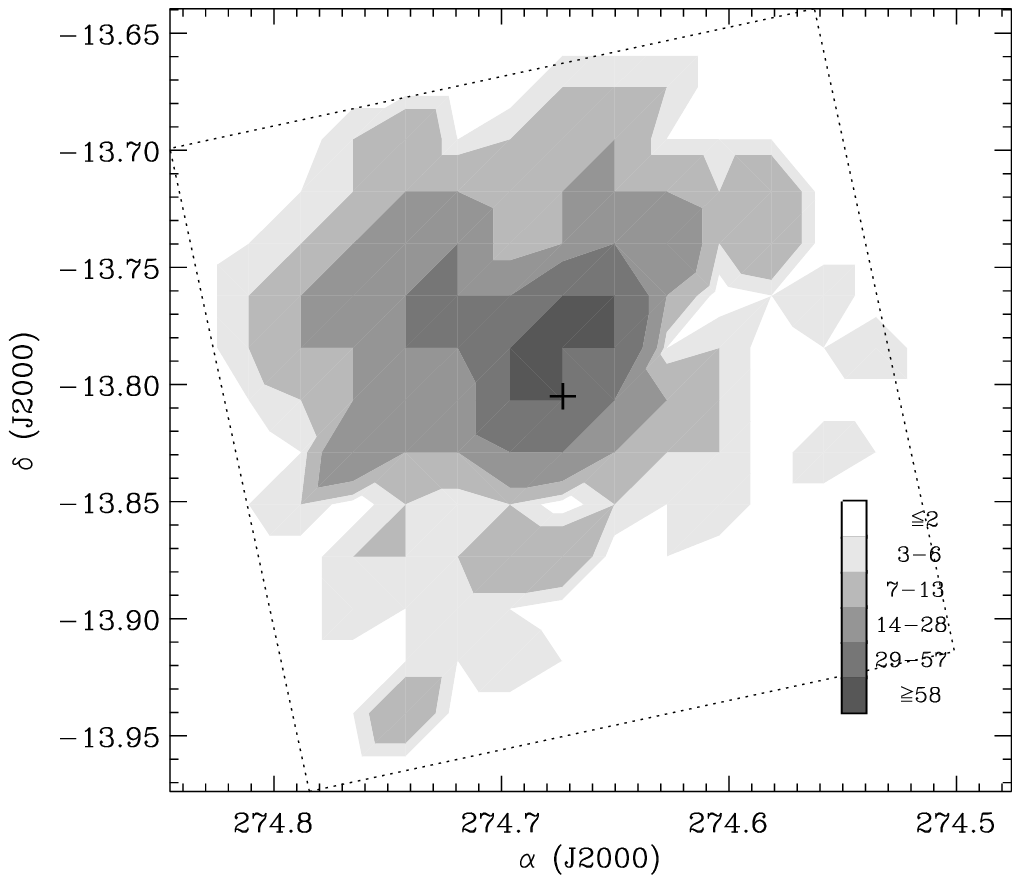}
	\includegraphics[width=6.7cm]{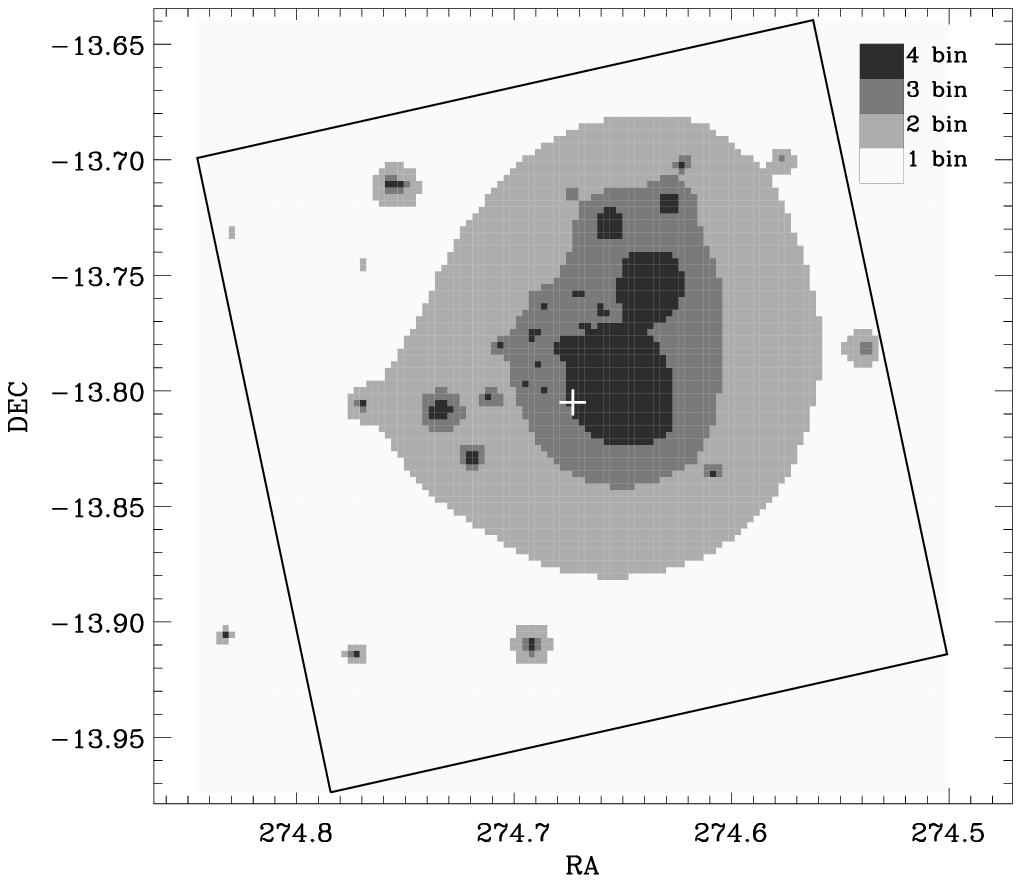}
	\includegraphics[width=6.7cm]{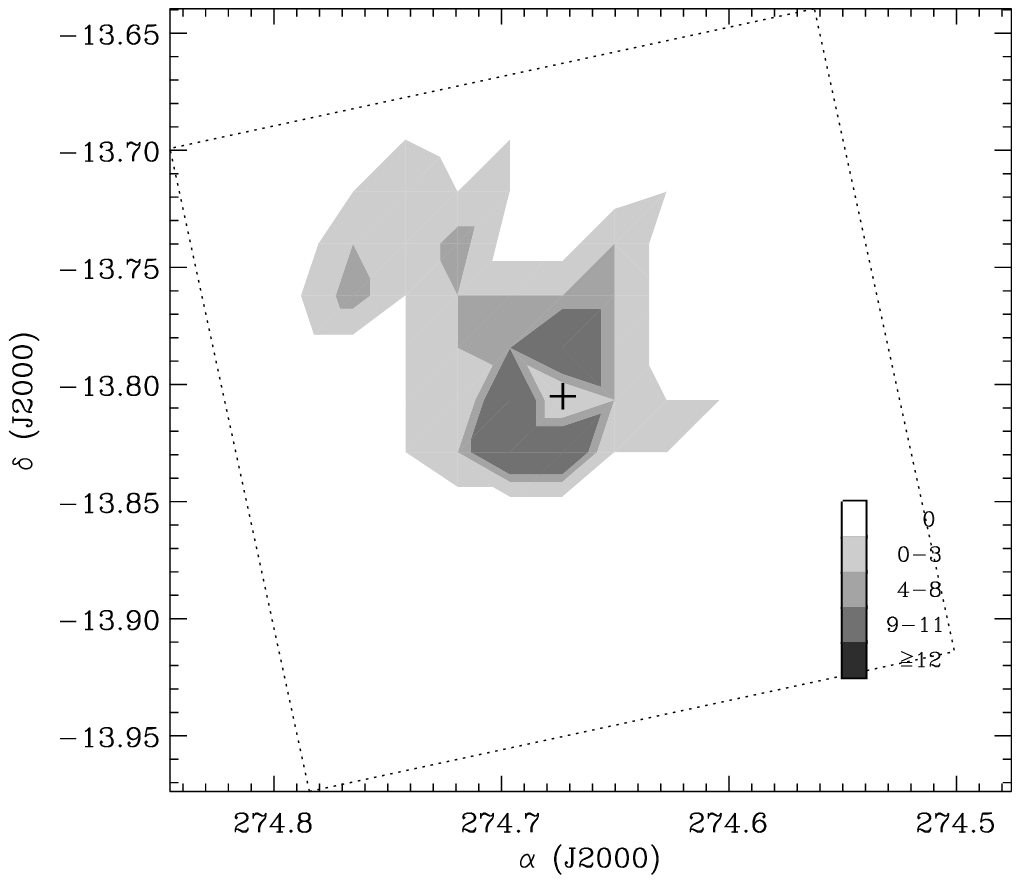}
	\includegraphics[width=6.7cm]{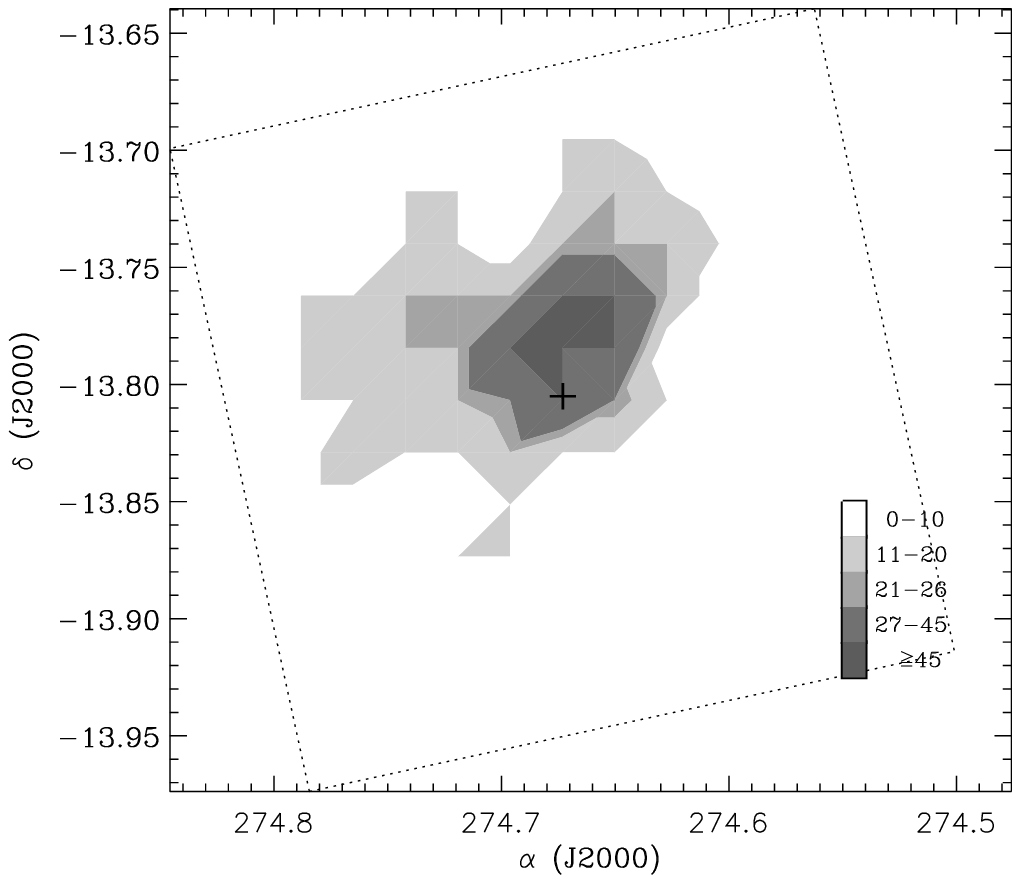}
	\caption{The upper left panel shows the spatial distribution of candidate cluster members. The dotted square delimits the ACIS FOV. The upper right panel shows the intensity map of the radiation emitted by the massive members of the cluster. The different shades correspond to the four bins in Fig. \ref{isto}. The lower left panel shows the spatial distribution of candidate members with disk; the lower right panel that of candidate members without disk. In all the panels, the cross marks the center of the ACIF FOV}
	\label{maps}
	\end{figure*}

The panels in Fig. \ref{maps} allow to compare directly the spatial distribution of the cluster members and the intensity of the radiation emitted by massive members. The cluster members are not equally distributed in the ACIS FOV, with an empty region at South-West. The intensity map shown in the upper right panel of Fig. \ref{maps} is more concentrated that the members density, reflecting the central concentration of massive members. \par
	The two lower panels of Fig. \ref{maps} show the spatial distributions of disk-bearing and disk-less members. Note that the latter stars have a more symmetric distribution; while the former are more present where the UV flux is low, and also their distribution present an ``hole'' in the center of the field, where the UV flux is lower. Moreover, the region corresponding to the highest fluxes of Fig. \ref{isto} have a radius of 0.6 parsecs, the distance from massive stars at which \citet{Bal07} found a drop in disk frequency in NGC~2244.
	
	The upper right panel of Fig. \ref{maps} also allows a more easy confrontation of our result with that of \citet{Bal07}. As explained, these authors studied the spatial variation of the disk frequency respect to the positions of massive stars in the young cluster NGC~2244. They have not found a correlation between the two distributions, but observed a significant drop in disk frequency for distance smaller than 0.5 pc from O stars. The central regions of the intensity map in Fig. \ref{maps}, corresponding to the highest fluxes in Fig. \ref{isto}, has a radius equal to about 0.6 parsec, so the decrease of the disk frequency in the last bin in Fig. \ref{isto} is in agreement with the finding of \citet{Bal07}. \par

To have an estimation of the extension of NGC~6611, even if it has not a symmetric spatial distribution, we use the 2-parameters density profile of \citet{Ki66}:
	
	\begin{equation}
	\sigma(r)=\frac{\sigma_0}{1+\left(r/r_{core} \right)^2}
	\label{kingeq}
	\end{equation}
	
	where $\sigma_0$ is the central cluster density, $r$ is the distance of stars from cluster center and $r_{core}$ is the core radius. We calculated from the observed radial density profile: $\sigma_0 = 149$ $\pm 8 N_{stars} pc^{-1}$ and $r_{core} = 1.39 \pm 0.08 pc$, in agreement with the results obtained in GPM07. 
	

\section{Summary}
\label{thatsallfolks}

	In this paper, we analyze Spitzer/IRAC data of the young open cluster NGC~6611, in the Eagle Nebula, in a large field wide approximately $30^{\prime}\times30^{\prime}$. In a previous paper, we have already selected the members of this cluster using our published BVIJHK and X-ray multi band catalog. The obtained list of cluster members, both with and without disk, is used to verify that the former are more frequent at larger distances from the massive members of the cluster. \par
In this work, in order to select new members with disk, we use and compare two different disk diagnostics: the IRAC color-color diagram, that use all the IRAC bands simultaneously, and four suitable reddening-free color indices ($Q$ indices), each defined to detect the excesses in a specific IRAC band. \par
	Using the IRAC color-color diagram, we identify 182 Young Stellar Objects (147 Class~II; 13 Class~I and 22 with colors compatible with both classifications). This diagnostic is very efficient to select largely embedded young sources, but it can be used only for stars with good photometry in all IRAC bands. We discuss how significantly this affects the selection of faint YSOs. With the use of $Q$ indices we partially overcome this problem, since they allow us to select stars with excesses in each IRAC band taken alone; with this diagnostic, we select 174 YSOs with excesses in IRAC bands. Among them, 66 stars cannot be detected with the IRAC color-color plane, since they are not well measured in all the IRAC bands.\par
Combining the outcome of both diagnostics, we identify 146 new cluster members with disk with respect to our previous work, which was based on 2MASS photometry alone (116 from the IRAC color-color diagram and 30 from $Q$ indices). At the end, our catalog includes a total of 474 candidate Pre-Main Sequence members with disk and 790 without disk (the latter only in the central $17^{\prime}\times17^{\prime}$ field, where we found 118 X-ray sources with disk). \par
	Comparing all the $Q$ indices defined with 2MASS and IRAC bands (with the exception of [8.0] band), we claim that they are all sensitive to the emission from the same physical region of the disk, namely the inner rim at the dusts sublimation radius. These indices will be efficiently used, then, to select Class~II stars that are not largely embedded (i.e. for which it is possible to detect the photospheric emission) and whose inner disk is not evacuated. The embedded YSOs associated to the cluster can be selected only thank to the IRAC color-color diagram, and not with the $Q$ indices.\par
We discuss evidence suggesting that star formation activity is ongoing in the outer region of the Eagle Nebula, and not only in the center of NGC~6611 as showed by previous works. For example, the diagnostic based on IRAC color-color diagram allow us to identify a probable new star formation site, rich of Class~I and embedded Class~II objects, at North with respect to the center of the cluster. \par
	 In the central $17^{\prime}\times17^{\prime}$ field we find an average disk fraction equal to $24\% \pm 2\%$. In this field, the disk frequency also has a strong dependence on the incident radiation emitted by massive members, varying from $31\% \pm 4\%$ to $16\% \pm 3\%$ across the whole range of values of incident flux. This is an evidence of the influence of massive stars radiation on the evolution of circumstellar disks and star formation process, as already shown in GPM07.
	

\begin{acknowledgements}
We thank the anonymous referee for its useful comments that allowed us to improve our manuscript. We acknowledge financial support from the contract PRIN-INAF. This publication makes use of Spitzer's GLIMPSE survey data. 
\end{acknowledgements}

\addcontentsline{toc}{section}{\bf Bibliografia}
\bibliographystyle{apj}
\bibliography{biblio}

\end{document}